# Insights, opportunities and challenges provided by large cell atlases

Martin Hemberg[1,2,*], Federico Marini[3,4,*], Shila Ghazanfar[5,6,7,*], Ahmad Al Ajami[8,9,10], Najla Abassi[3,4], Benedict Anchang[11], Bérénice A. Benayoun[12,13,14], Yue Cao[5,6,7,15], Ken Chen[16], Yesid Cuesta-Astroz[17,18], Zach DeBruine[19], Calliope A. Dendrou[20], Iwijn De Vlaminck[21], Katharina Imkeller[8,9,10], Ilya Korsunsky[22,2,23], Alex R. Lederer[24], Pieter Meysman[25], Clint Miller[26], Kerry Mullan[25], Uwe Ohler[27], Nikolaos Patikas[1,2], Jonas Schuck[8,9,10], Jacqueline HY Siu[20], Timothy J. Triche, Jr.[28], Alex Tsankov[29], Sander W. van der Laan[30], Masanao Yajima[31], Jean Yang[5,6,7,15], Fabio Zanini[33], Ivana Jelic[34]

1 The Gene Lay Institute of Immunology and Inflammation, Brigham and Women's Hospital, Massachusetts General Hospital
2 Harvard Medical School, Boston, MA, USA
3 Institute of Medical Biostatistics, Epidemiology and Informatics (IMBEI), University Medical Center Mainz, Mainz, Germany
4 Research Center for Immunotherapy (FZI), Mainz, Germany
5 School of Mathematics and Statistics, Faculty of Science, University of Sydney, NSW 2006, NSW, Australia
6 Sydney Precision Data Science Centre, University of Sydney, Sydney, NSW 2006, Australia
7 Charles Perkins Centre, University of Sydney, Sydney, NSW 2006, Australia
8 Goethe University, Neurological Institute / Edinger Institute, University Hospital Frankfurt, Frankfurt am Main, Germany
9 Goethe University, Frankfurt Cancer Institute, Frankfurt am Main, Germany
10 University Cancer Center (UCT), Frankfurt am Main, Germany
11 National Institute of Environmental Health Sciences
12 Leonard Davis School of Gerontology, University of Southern California, Los Angeles, CA 90089, USA.
13 Molecular and Computational Biology Department, USC Dornsife College of Letters, Arts and Sciences, Los Angeles, CA 90089, USA.
14 Biochemistry and Molecular Medicine Department, USC Keck School of Medicine, Los Angeles, CA 90089, USA.
15 Laboratory of Data Discovery for Health Limited (D24H), Science Park, Hong Kong SAR, China
16 The University of Texas MD Anderson Cancer Center
17 Escuela de Microbiología, Universidad de Antioquia, Ciudad Universitaria Calle 67 No 12 53-108, Medellín, Colombia.
18 Instituto Colombiano de Medicina Tropical, Universidad CES, 055413 Sabaneta, Colombia
19 School of Computing, Grand Valley State University, Allendale, MI 49401, USA
20 Kennedy Institute of Rheumatology, Nuffield Department of Orthopaedics, Rheumatology and Musculoskeletal Sciences, University of Oxford, Oxford, UK
21 Cornell University
22 Brigham and Women's Hospital, Divisions of Genetics and Rheumatology, Boston, MA, USA
23 Broad Institute of Massachusetts Institute of Technology (MIT) and Harvard, Cambridge, MA, USA
24 Laboratory of Brain Development and Biological Data Science, Brain Mind Institute, School of Life Sciences, École Polytechnique Fédérale de Lausanne (EPFL), 1015 Lausanne, Switzerland
25 Adrem data lab, Department of Computer Science, University of Antwerp, Antwerp, Belgium
26 The Rector and Visitors of the University of Virginia
27 Max Delbruck Center for Molecular Medicine
28 Department of Epigenetics, Van Andel Institute, Grand Rapids, MI, US
29 Icahn School of Medicine at Mount Sinai
30 University Medical Center Utrecht
31 Boston University
32 School of Mathematics and Statistics, Faculty of Science, University of Sydney, NSW 2006, NSW, Australia
33 University of New South Wales
34 Chan Zuckerberg Initiative
* = These authors contributed equally

*Correspondence should be addressed to* mhemberg@bwh.harvard.edu, marinif@uni-mainz.de, shila.ghazanfar@sydney.edu.au, ijelic@chanzuckerberg.com

*All authors participated in the series of meetings organized by CZI Single-Cell Biology program meetings*

## Abstract

The field of single-cell biology is growing rapidly and is generating large amounts of data from a variety of species, disease conditions, tissues, and organs. Coordinated efforts such as CZI CELLxGENE, HuBMAP, Broad Institute Single Cell Portal, and DISCO, allow researchers to access large volumes of curated datasets. Although the majority of the data is from scRNAseq experiments, a wide range of other modalities are represented as well. These resources have created an opportunity to build and expand the computational biology ecosystem to develop tools necessary for data reuse, and for extracting novel biological insights. Here, we highlight achievements made so far, areas where further development is needed, and specific challenges that need to be overcome.

## Introduction

Technological advances have enabled generation and collection of large volumes of data at the single-cell resolution (Svensson, da Veiga Beltrame, and Pachter 2020). For the most part, this is done by individual research groups, and to make these datasets more useful to the community they need to be assembled into cell atlases. In the context of single-cell technologies an atlas is a large collection of datasets that have been curated, and made accessible through a web portal. In addition to making it easier to find datasets, atlases provide a coherent pipeline for data ingestion and processing, ensuring that datasets can be combined and leveraged to provide novel biological insights.

Today, there are thousands of single-cell datasets available, and building an atlas is a resource-intensive endeavor, requiring a large team of biologists and data scientists. Moreover, substantial infrastructure is needed, and to be useful to the community it must be sustained and updated over time. Hence, cell atlases are backed by large organizations, or supported by large-scale projects (Table 1).

Table 1: Summary statistics for selected single cell atlases

| Name | #cells | #donors | Year started | Organization | #species | URL | Reference |
|---|---|---|---|---|---|---|---|
| CZ CELLxGENE Discover | 84.3M | 3.8k | 2022 (launch) | CZI | 7 | https://cellxgene.cziscience.com/ | 10.1101/2023.10.30.563174 (CZI Single-Cell Biology Program et al. 2023a) |
| Single Cell Portal | 43.2M | Not reported | 2020 (launch) | Broad Institute | 17 | https://singlecell.broadinstitute.org/single_cell | 10.1101/2023.07.13.548886 (Tarhan et al. 2023) |
| Single Cell Expression Atlas | 10.5M | Not reported | 2018 (launch) | EMBL-EBI | 21 | https://www.ebi.ac.uk/gxa/sc/home | 10.1093/nar/gkab1030 (Moreno et al. 2022) |
| Human BioMolecular Atlas Program (HuBMAP) | Not reported | 214 | 2018 (launch) | NIH | 1 | https://hubmapconsortium.org/ | 10.1038/s41556-023-01194-w (Jain et al. 2023) |
| Human Cell Atlas (HCA) | 58.5M | 8.6k | 2016 (launch) | HCA | 1 | https://data.humancellatlas.org/ | 10.7554/eLife.27041 (Regev et al. 2017) |
| Allen Brain Cell (ABC) Atlas | 4.0M | Not reported | 2023 (publication) | Allen Institute | 1 | https://portal.brain-map.org/atlases-and-data/bkp/abc-atlas | |
| Tumor Immune Single Cell Hub 2 (TISCH2) | 6.3M | Not reported | 2023 (publication) | Shanghai Putuo District People's Hospital | 2 | http://tisch.comp-genomics.org/ | 10.1093/nar/gkac959 (Han et al. 2023) |
| Deeply Integrated Single-Cell Omics (DISCO) | 18M | Not reported | 2022 (publication) | Singapore Immunology Network | 1 | https://www.immunesinglecell.org | 10.1093/nar/gkab1020 (Li et al. 2022) |
| Single Cell Atlas | 70M | 125 | 2024 (publication) | SCA Consortium | 1 | https://www.singlecellatlas.org | 10.1186/s13059-024-03246-2 (Pan et al. 2024) |

**Table 1: A selection of cell atlases currently available.** The number of cells corresponds to the approximate number of cells with transcriptomics readout at the time of writing.

Together, we have been involved in the Chan Zuckerberg Initiative's effort to expand the ecosystem of computational methods that can support and exploit cell atlases in the context of the Data Insights program. Here, we present our shared experiences and we discuss some of the issues involved in building and using a cell atlas (**Fig. 1**). We then go on to explore the possibilities enabled by cell atlases as well as the challenges that the community faces going forward.

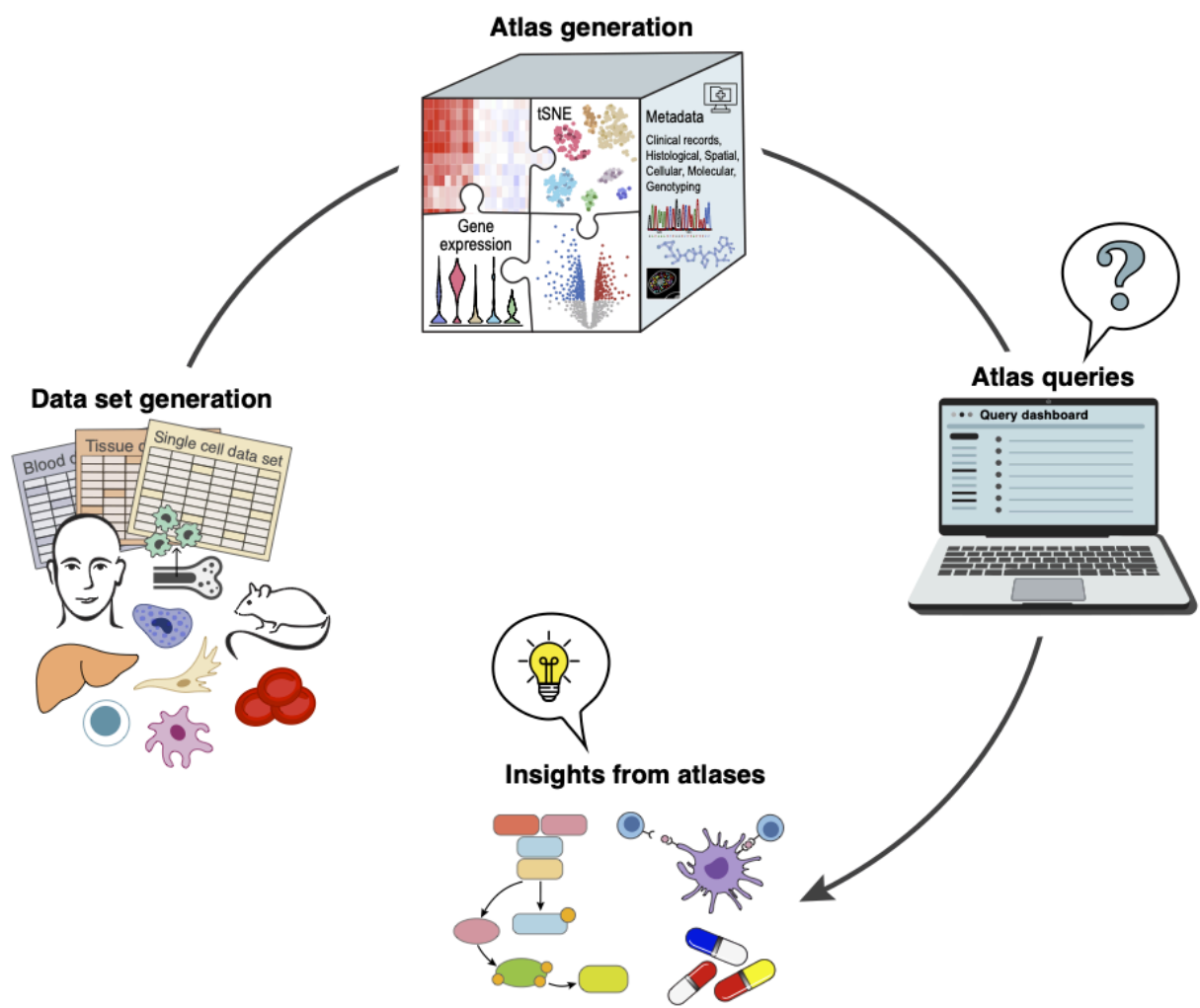

**Figure 1:** Cell atlases ingest data from a wide range of labs based on specific criteria, e.g. species, disease, tissue. Data is processed in a coherent manner and made available through a portal. The portal enables a wide range of queries to either download or interrogate multiple datasets. On their own or in combination with additional experiments, these queries can result in new findings.

# Data Ingestion, Access and Representation

A central goal for any scientific resource is to make sure that it adheres to the FAIR principles (Wilkinson et al. 2016), i.e. ensuring that data is findable, accessible, interoperable, and reusable. By serving as central repositories, cell atlases make it easier to find and access data. By making sure that data is uniformly processed and adheres to standard formats, it also becomes interoperable and reusable. Although straightforward in principle, the scale and complexity of a cell atlas make it difficult to achieve these goals.

## Data pre-processing

For sequencing data to be useful, one must have access to the underlying reads, typically stored in fastq format (Cock et al. 2010). In addition to the raw data, the various levels of processed data and metadata must also be ingested. The first step for the cell atlas is to carry out quality controls to ensure the integrity of the data. Thus, preprocessing is a key step that is often poorly documented, and difficult to reproduce due to the use of different versions of software packages. Although preprocessing can improve the internal consistency, it does not prevent the emergence of discrepancies within and across atlases. A particular concern for cell atlases is batch effects, technical artifacts that emerge due to differences in how the data was obtained and processed. Although batch effects can be reduced, they cannot be eliminated altogether. Fortunately, it is possible to detect and correct batch effects post-hoc, provided that detailed information about the processing is available. As no repository will be able to span all conditions, populations, organisms, cells, and modalities of interest, maintaining this possibility is a key requirement for enabling meta-analyses.

## Data accessibility, interoperability, and reusability

By providing a portal where users can search for data, cell atlases greatly facilitate finding datasets. Depending on the use case, different levels of processing will be desired. Only providing the raw data is not sufficient, and various levels of processing will be required by most users. However, tools for indexing, metadata standardization, and cross-cohort queries are in their infancy, and this limits the ability of users to identify suitable datasets (Fischer et al. 2021; CZI Single-Cell Biology Program et al. 2023a). Although it is possible to find and access individual datasets through a web browser, anyone interested in analyzing a large number of datasets needs to have both programming skills and sufficient computational resources. This creates a barrier for many users, and an important area of research is to develop computational tools to facilitate access to large cell atlases. Another key aspect is to provide APIs for those users who are developing code for accessing the cell atlas. This includes adhering to standardized file formats as well as catering for multiple programming languages. At the time of writing, both R and python are widely used, and cell atlases need to support both to be useful. As with many other aspects of a cell atlas, these resources need to be updated over time as the computational eco-system and use cases evolve.

## Metadata and ontologies

Metadata is crucial for researchers interested in re-analyzing existing datasets. This issue has long been recognized, and in the past, there have been community efforts such as MIAME (Brazma et al. 2001) to set common standards. For scRNAseq, metadata can be divided into three categories: sample, gene, and cell. Sample metadata includes information about donor, time of collection, storage, experimental processing, computational processing, etc. Gene metadata is relatively straightforward, at least for model organisms, where various annotations are mature and stable. The most important aspect of cell metadata is its annotation, and this requires mapping data to a cell type ontology (Diehl et al. 2016). An example of a metadata scheme developed for single-cell analysis is matrix and metadata standards (MAMS, (Wang et al. 2023)). Although reporting and adhering to technical standards is important, it is essential to couple this to the establishment of a culture where data generators recognize their responsibility in providing complete metadata.

Ontologies allow formal and structured operations to be carried out and they are thus essential to contextualize the resource and facilitate interpretations. In particular, they enable automated processing and application of ML/AI methods. The most commonly used ontology is one for cell types/states (Osumi-Sutherland et al. 2021), but there are other ones to consider. Each ontology captures one specific perspective, e.g. development, anatomy, disease, and hence it is valuable to have multiple, overlapping ones. For other modalities expanded ontologies will be required. Ontologies are also required at the sample level. It is particularly important with regards to disease samples where one must ensure that relevant clinical information is represented. For international projects, this can become problematic, as different healthcare systems may have different terminologies and criteria.

# Extracting the most out of a cell atlas

The immediate use of cell atlases is to provide a global overview of cell types and cell states for a given tissue, disease, organism, or condition. An inventory of the building blocks is of great scientific value, and once generated, it serves as a springboard to address further biological questions. The challenge for the research community is that there is a breadth of needs for accessing a cell atlas. For some researchers, it may be enough to access a web server that displays gene expression and cell clusters, while bespoke analysis access may require downloading cell atlases to a local computer or server.

## Data representation and subsampling

The typical workflow when using a cell atlas requires a user to first identify and download the relevant datasets. Given their size, already exceeding 1Tb using standard data structures, this can require significant bandwidth and be prohibitive to many research labs without high-memory computing resources. For most users, working with this data requires out-of-core processing, high-performance computing, and significant effort in data wrangling (Ruiter et al. 2023). Consequently, there is an urgent need for methods for handling streaming data as well as lossless compression algorithms for single-cell data that significantly reduce memory footprint without compromising computational performance. New data structures for R and R/C++/Python can reduce memory footprint by up to 10-fold over standard sparse matrices with minimal cost to compute.

One algorithmic approach for dealing with large datasets is to subsample. Subsampling can ease computation over diverse datasets and help reduce bias of highly represented signals, but may also compromise the unprecedented modeling power that comes with a dataset of this size. Issues include racial and gender bias in samples (Vaidya et al. 2024), over-representation of specific cell types (Truong et al. 2023), opportunistic collection of rare samples (Haniffa et al. 2021), but the best approach depends on the scientific goals of the study. Simple random subsampling does not address signal representation and can miss rare subpopulations (Song et al. 2022; Hie et al. 2019), and it is desirable to balance the tradeoff between representing the true proportions of cell types and full extent of cell identities of rare cell types. As cell atlases become more diverse across tissues and patient donors, subsampling will become more attractive. However, such summaries may impair our ability to tease out subtle and biologically relevant signals that only come with massive statistical power offered by large sample sizes. We also envision that latent space representations will be highly useful for providing compact representation, but further research is required to understand the accuracy and limitations of these approaches.

## Data integration and meta analysis

To relate cell atlases with each other or with additional single-cell datasets, the research community relies on data integration tools (Luecken et al. 2022). These tools aim to identify joint low-dimensional representations of all the input data, with which further downstream analysis can be performed, e.g. joint clustering, cell type classification, and differential abundance testing. In modern single-cell analysis, this data integration is often key for harmonizing distinct datasets, and it serves as an initial step for meta-analysis of single-cell datasets. Appropriate meta-analyses will not only have to consider corrected cell type labels, but also other statistical factors that remain challenging to deal with. Issues such as confounding factors, nested structure of single cells measured within biological samples (i.e. repeated measures), and understanding hidden sources of variability all may be relevant. Ideally, cells are curated with sample-level information such as donor identity, sex, age, tissue, organism, developmental stage, technology, and disease. Some of these confounders are technical and performing data integration over them will remove noise and increase salience of biological signals (Emmanúel Antonsson and Melsted 2024). Others are biologically driven, and thus data integration would enable researchers to compare analogous cell states across tissue, diseases, and organisms. A useful data integration algorithm must account for all of these sources of information and allow users to retain or remove specific variation relevant to their analyses. For instance, a comparison of T cells across tissues may emphasize tissue-specific differences (i.e. do not harmonize over tissue) to explain differential heritability, while others look to nominate for shared effector phenotypes across diseases (i.e. do harmonize over tissue) for basket clinical trials (Mullan et al. 2023). These scenarios highlight that data integration is not a static tool to be applied prior to further analysis, but rather needs to be adaptable to the research question at hand. This introduces a new computational challenge, as current integration algorithms are designed with the idea that they are run once per analysis, and therefore are not necessarily optimized to an online dynamic query setting. Successful data integration algorithms must address these emerging complexities and they must (1) scale to 10,000s of confounder levels, (2) account for all sources of technical and biological variation in a way that lets users select which to account for, (3) perform consistently across a wide variety of cell atlas queries, (4) be fast and flexible enough to integrate diverse queries on the fly, (5) provide views of their impact on data distortion and signal degradation in a manner that is easy for the user to interpret.

## Building cell atlases in context

An important use of cell atlases, beyond defining cell types and corresponding markers, is to explore how cellular and transcriptional landscapes are impacted by specific disease/functional decline (e.g. disease states, reduced function), physiological/biological factors (e.g. age, sex/gender, ethnic/genetic background), or treatments (e.g. response to a drug) (Thomas et al. 2023; Ren et al. 2021; Chu et al. 2023). Indeed, important insights can be achieved by analyzing cell atlases in a context-aware fashion, both comparing how biologically relevant inputs can lead to changes in (i) cell composition (Phipson et al. 2022; Cao et al. 2019; Miller et al. 2021) and (ii) cell-type specific gene expression. Importantly, even for single-cell atlases, biological replicates should consist of samples procured from independent biological sources/individuals, and not just independent cells from the same sample (Squair et al. 2021a; Crowell et al. 2020a; Dann et al. 2022). Thus, a key feature of context-aware cell atlases is the inclusion of sufficient independent biological samples across conditions to account for inter-individual variability. This requires sufficient numbers of true biological replicates, similar to bulk approaches (Schurch et al. 2016). For the interpretation of context-dependent cell atlases, it is crucial to consider the need for approaches that limit the high false positive rates in single-cell differential analyses (e.g., considering the potential of pseudobulking approaches per identified cell type/state to avoid underestimation of true biological variability (Squair et al. 2021b; Crowell et al. 2020b)). As an example, in the study of menopause and its potential molecular drivers, a context-aware atlas should include sufficient samples covering both pre-menopausal and post-menopausal states, in tissues most relevant to the condition (*e.g.* ovary, pituitary gland, hypothalamus) (Singh and Benayoun 2023).

## Benchmarking and development of novel methods

There is currently a rapidly expanding ecosystem of computational tools in the single-cell field, and for most problems, there is more than one method available. To help researchers decide what tool to choose, benchmarking studies are essential, and multiple benchmarking papers are published every month. Several challenges exist in the current single cell benchmarking field and guidelines on best practices are needed. It needs to be clear what are the evaluation metrics and although many details will depend on the specific topic there are overarching trends (Cao et al. 2023). One of the main challenges when comparing methods is that for most problems we do not have an independent ground truth. Hence, evaluating the performance will involve some degree of subjectivity. One way of circumventing this challenge is to use simulations to create synthetic data. However, most synthetic datasets are unable to capture the full spectrum of complexities found in real datasets, and more work is required to build on recent developments (Song et al. 2024). Comparisons using real datasets require curation, a process that can be time-consuming and requires substantial domain knowledge. Moreover, it may further entrench any value of certain algorithms through the process in performing the curation (e.g. clustering algorithms). As such, there is tremendous value in datasets that can be considered as a 'gold standard' by virtue of orthogonal means. Carefully curated cell atlases can serve an important role here by being commonly used for benchmarking specific analytical tasks.

Today, most analysis tools and strategies are designed within the context of a smaller number of datasets or total cells. Many methods that are commonly used may not scale well to tens of millions of cells and thousands of conditions, and consequently, there is a need to increase computational and algorithmic efficiency. This is likely to involve various types of approximations and lossy compression to achieve the desired speed-up and reduction in memory footprint. An example is the use of strategies like mini-batch to speed up the estimation of k-means clustering (Hicks et al. 2021) without compromising the quality of results.

Alongside improvements to existing analytical strategies, a wider reformulation of these methods to address new biological questions is needed to fully leverage datasets spanning tissues, species, and organismal age. One such example of an established method that requires novel frameworks to be applicable to new types of data is RNA velocity. A manifold-constrained and biologically-tailored velocity, as opposed to general-purpose tools, could be designed to statistically compare estimates in the case-control setting (Aivazidis et al. 2023; Lederer et al. 2024). With such a modular method, a diseased or otherwise perturbed sample could be used to investigate whether subtle disruptions of the RNA velocity vectors indicate an effect of a particular perturbation.

Recently, there have been tremendous advances in AI, in particular in applications related to natural language processing, protein structure prediction, and image processing. What these methods have in common is that they rely on large datasets for training, and consequently cell atlases will help their advance. Recently, several groups have developed foundational models based on cell atlases, e.g. Geneformer and scGPT (Cui et al. 2024; Theodoris et al. 2023), demonstrating how an atlas can be leveraged. However, understanding how to adopt advances in deep learning remains an open question.

## Using cell atlases for biomedical research

Perhaps the most important application underpinning efforts to build cell atlases is the notion that they can help accelerate biomedical research to help manage and cure disease (Rood et al. 2022a). Below, we discuss some of the areas where cell atlases will provide key resources.

From large-scale genome-wide association studies (GWAS), thousands of genetic loci have been identified that infer risk of disease or influence human traits. While these studies have yielded great and unexpected insights into complex and common diseases, they also reveal a yawning knowledge gap. For instance, for 50% of the risk loci identified for coronary artery disease (CAD) it is unclear which gene(s) and which cell(s), and therefore which molecular and cellular pathways may be involved (Chen and Schunkert 2021). From studying CAD associated loci it is clear that gene regulatory effects are context-dependent, and that genetic effects can be condition-specific in terms of effect, direction, and magnitude (Aherrahrou et al. 2023). In other words biological sex, environmental factors (e.g. smoking), and disease influence genotypic effects and change cellular gene expression and responses. The analysis of cell type and condition-specific genetic effects on cellular molecular traits quantitative trait locus (molQTL) analysis will provide critical insights. These efforts could enable cell type and cell state-specific colocalization of molQTL and GWAS signals to interpret the regulatory mechanisms for complex diseases and traits. Cell atlases that combine genetic information with cell molecular profiles will be a key resource in unraveling such complex regulatory effects.

Cell atlases can also facilitate therapeutic target discovery, e.g. by predicting disease-relevant cell states by identifying gene signatures along a trajectory from healthy to disease (Van de Sande et al. 2023). For example, muscle cells downregulate their contractile markers and become mesenchymal stem-like cells, before adopting specialized cell states (Grootaert and Bennett 2021). There are now automated pipelines (e.g., scDrug, Drug2Cell (Hsieh et al. 2023; Kanemaru et al. 2023)) that take as input a cell-by-gene matrix of protein-coding genes and leveraging the full compendium of drug-gene interactions (e.g., DGIdb (Cannon et al. 2024)), cell perturbations (e.g. LINCS L1000 (Subramanian et al. 2017)),  FDA-approved molecules and biologics (e.g. DrugBank (Knox et al. 2024)), or active ligands (e.g. ChEMBL (Zdrazil et al. 2024)) to prioritize potential drug target genes. Importantly, atlases could also be used to predict drug responses or unwanted side effects (e.g. scDR (Lei et al. 2023)) by querying identified targets in public databases (e.g. SIDER (Kuhn et al. 2016)). Such a combination of cell atlas and therapeutic databases would enable the combination of genetic epidemiology, in particular, causal inference through Mendelian randomization, with single-cell biology, resulting in effective identification of druggable targets or surrogate markers of disease.

The large collections of data presented in cell atlases require appropriate tools and frameworks that enable efficient exploration. Only in this way can they fulfill their potential in assisting a wide spectrum of researchers to generate novel hypotheses, faithfully representing the results, and communicating the findings with the community. Several platforms and interfaces that aim to simplify the extraction of insight from such datasets have proliferated over the last decade, including for example the CELLxGENE tool (CZI Single-Cell Biology Program et al. 2023b), the Bioconductor iSEE package (Rue-Albrecht et al. 2018), and the browsers included in the Broad Single Cell Portal or the Single Cell Expression Atlas (Moreno et al. 2021). The spectrum of operations covered by such tools enables a powerful, in-depth exploration, possibly blending different views and representations of these large corpora of data, and linking out to other existing databases or relevant resources.

## Beyond atlases of dissociated single-cell transcriptomes

Single-cell RNA-seq was the first high throughput method that allows for a high-plex characterization of individual cells, and thus it has been the most widely used approach (Svensson, Vento-Tormo, and Teichmann 2018; Svensson, da Veiga Beltrame, and Pachter 2020). However, there are numerous other single-cell technologies under active development, and we foresee that over the coming years cell atlases will see an influx of other modalities (Zhu, Preissl, and Ren 2020). These include TCR and BCR sequencing, ATAC-seq, and long-read sequencing. Although this is likely to be hugely beneficial to researchers, it also involves several different challenges. This starts with the organizations supporting the cell atlas which must develop standards and protocols for how to process and curate other modalities (Heumos et al. 2023). Ensuring that different modalities can be combined for joint analyses is key, but it will present some major challenges, e.g. developing pre-processing pipelines, ontologies, and determining what metadata to include.

Several assays have been developed for measuring other aspects of the cellular state in single cells, e.g. DNA methylation, accessible chromatin (ATAC-seq), and transcription factor binding (scCUT&Tag), and an active area of research is to apply them to the same cell for multiomics profiling. This will provide numerous opportunities, e.g. by combining ATAC-seq and RNA-seq data, we are likely to improve our ability to infer gene regulatory networks (Zhang et al. 2024).

Perhaps the most important direction of new technologies is toward spatial methods, primarily for transcriptomics and proteomics, but other modalities are likely to follow. Spatial data brings numerous challenges along with great potential for additional insights (Williams et al. 2022). One challenge is in visualizing the data, and here a user should be able to seamlessly toggle between gene expression space (typically a UMAP) and physical space. This representation is relatively straightforward for individual datasets, but for multiple samples it becomes much more challenging. There is also a need for further algorithmic development of methods and tools that can be used to mine the spatial data, e.g. finding subcellular patterns in cells associated with disease, type, and outcomes (Dries et al. 2021). As technologies evolve and are able to profile larger tissue sections at subcellular resolution, a particular challenge will be to develop methods that can bridge these length scales. Given that image analysis has been one of the areas where machine learning and AI methods have been most successful, it is clear that there are ample opportunities for drawing inspiration from these methods.

### Cell atlases outreach

At the moment, cell atlases are being built by scientists for other scientists (Quake 2022). However, given the potential implications to wider society and the substantial amounts of resources, much of which is coming from public funding, it is important that cell atlases can also cater to other audiences (HuBMAP Consortium 2019; Quake 2022). Beyond the core constituency of academic biomedical researchers, potential users include clinicians and researchers in biotech and pharmaceutical industries. However, we believe that the ambition should be to make the resource at least somewhat accessible to the general public, including patients, teachers, and students of all ages (Rood et al. 2022b). This has the potential of helping to educate the public on the advances and benefits of biomedical research, involving citizen scientists, and inspire the next generation of scientists to ensure that others will be able to build on the work.

## Conclusions and outlook

Here, we have outlined some of the challenges and opportunities brought along by cell atlases. We foresee that over the coming years, the existing atlases will continue to grow and that additional, more specialized collections will emerge. Having multiple atlases is likely to be beneficial to the field as different policies for curation, representation, interaction, and use cases. One analogy is gene annotations, where resources such as Ensembl, Refseq, and Gencode continue to be used in parallel. Depending on the specific need, one of these overlapping and complementary resources will be the most useful. In parallel, we expect substantial advances in computational methods that can effectively work with atlas-scale datasets to extract new insights.

Cell atlases have been enabled by technological advances, and we envision that continued innovation will govern how cell atlases evolve. As costs fall, we also expect that atlases will be broadened. To realize the potential not just for biomedical research, but for other aspects of biology, we need better coverage of human populations across the age spectra and for multiple disease states. Moreover, additional species are needed. Most likely, within the next few years advancements in the fields of proteomics and metabolomics will enable the profiling of large numbers of single cells via these modalities, unlocking more accurate modeling of metabolism, signaling and cell-cell communication.

The availability of large-scale data resources contributes to the democratization of science. In fact, we have reached a point where many projects are run using only public data. Single-cell data is information dense, and there is little chance that the lab that generated the data has the capacity to make all the discoveries that are possible, in particular those that are only possible by combining with other datasets. Enabling the efficient use and reuse of complex and multiple datasets allows our scientific community to increase the pace of scientific discoveries.

In summary, we have described the development and utilization of cell atlases, which are comprehensive maps of cell types and states generated through the integration of large volumes of single-cell data. We detailed challenges and opportunities in building, maintaining, and utilizing these atlases (**Figure 2**), emphasizing the importance of data standardization, accessibility, and computational tools for extracting meaningful biological insights, ultimately aiming to facilitate cross-tissue, cross-condition, and cross-species studies in the field of single-cell biology.

**Atlas-era challenges / best practices**

**QC starting materials:**
- good annotation
- good metadata
- thoughtful queries

**Atlas potential:**
interesting biological questions my require new computational methods

**Constantly evolving:**
- field is evolving
- new modalities introduced
- requires atlases to change to keep up

**Figure 2: Key challenges, opportunities, and issues regarding cell atlases today.**

# Acknowledgements

We would like to thank Leslie Gaffney for assistance with the figures. This work was funded by the Chan Zuckerberg Initiative.

# Full Author List

| Author | Affiliation |
|---|---|
| Martin Hemberg | The Gene Lay Institute of Immunology and Inflammation, Brigham and Women's Hospital, Massachusetts General Hospital<br>Harvard Medical School |
| Federico Marini | Institute of Medical Biostatistics, Epidemiology and Informatics (IMBEI), University Medical Center Mainz, Mainz, Germany<br>Research Center for Immunotherapy (FZI), Mainz, Germany |
| Shila Ghazanfar | School of Mathematics and Statistics, Faculty of Science, University of Sydney, NSW 2006, NSW, Australia<br>Sydney Precision Data Science Centre, University of Sydney, Sydney, NSW 2006, Australia<br>Charles Perkins Centre, University of Sydney, Sydney, NSW 2006, Australia |
| Ahmad Al Ajami | Goethe University, Neurological Institute / Edinger Institute, University Hospital Frankfurt, Frankfurt am Main, Germany<br>Goethe University, Frankfurt Cancer Institute, Frankfurt am Main, Germany<br>University Cancer Center (UCT), Frankfurt am Main, Germany |
| Najla Abassi | Institute of Medical Biostatistics, Epidemiology and Informatics (IMBEI), University Medical Center Mainz, Mainz, Germany<br>Research Center for Immunotherapy (FZI), Mainz, Germany |
| Benedict Anchang | National Institute of Environmental Health Sciences |
| Bérénice A. Benayoun | Leonard Davis School of Gerontology, University of Southern California, Los Angeles, CA 90089, USA.<br>Molecular and Computational Biology Department, USC Dornsife College of Letters, Arts and Sciences, Los Angeles, CA 90089, USA.<br>Biochemistry and Molecular Medicine Department, USC Keck School of Medicine, Los Angeles, CA 90089, USA. |
| Yue Cao | School of Mathematics and Statistics, Faculty of Science, University of Sydney, NSW 2006, NSW, Australia<br>Sydney Precision Data Science Centre, University of Sydney, Sydney, NSW 2006, Australia<br>Charles Perkins Centre, University of Sydney, Sydney, NSW 2006, Australia<br>Laboratory of Data Discovery for Health Limited (D24H), Science Park, Hong Kong SAR, China |
| Ken Chen | The University of Texas MD Anderson Cancer Center |
| Yesid Cuesta-Astroz | Escuela de Microbiología, Universidad de Antioquia, Ciudad Universitaria Calle 67 No 12 53-108, Medellín, Colombia.<br>Instituto Colombiano de Medicina Tropical, Universidad CES, 055413 Sabaneta, Colombia |

| Name | Affiliation |
|---|---|
| Zach DeBruine | School of Computing, Grand Valley State University, Allendale, MI 49401, USA |
| Calliope A. Dendrou | Kennedy Institute of Rheumatology, Nuffield Department of Orthopaedics, Rheumatology and Musculoskeletal Sciences, University of Oxford, Oxford, UK |
| Iwijn De Vlaminck | Cornell University |
| Katharina Imkeller | Goethe University, Neurological Institute / Edinger Institute, University Hospital Frankfurt, Frankfurt am Main, Germany<br>Goethe University, Frankfurt Cancer Institute, Frankfurt am Main, Germany<br>University Cancer Center (UCT), Frankfurt am Main, Germany |
| Ilya Korsunsky | Brigham and Women's Hospital, Divisions of Genetics and Rheumatology, Boston, MA, USA<br>Harvard Medical School, Boston, MA, USA<br>Broad Institute of Massachusetts Institute of Technology (MIT) and Harvard, Cambridge, MA, USA |
| Alex R. Lederer | Laboratory of Brain Development and Biological Data Science, Brain Mind Institute, School of Life Sciences, École Polytechnique Fédérale de Lausanne (EPFL), 1015 Lausanne, Switzerland |
| Pieter Meysman | Adrem data lab, Department of Computer Science, University of Antwerp, Antwerp, Belgium |
| Clint Miller | The Rector and Visitors of the University of Virginia |
| Kerry Mullan | Adrem data lab, Department of Computer Science, University of Antwerp, Antwerp, Belgium |
| Uwe Ohler | Max Delbruck Center for Molecular Medicine |
| Nikolaos Patikas | The Gene Lay Institute of Immunology and Inflammation, Brigham and Women's Hospital, Massachusetts General Hospital<br>Harvard Medical School |
| Jonas Schuck | Goethe University, Neurological Institute / Edinger Institute, University Hospital Frankfurt, Frankfurt am Main, Germany<br>Goethe University, Frankfurt Cancer Institute, Frankfurt am Main, Germany<br>University Cancer Center (UCT), Frankfurt am Main, Germany |
| Jacqueline HY Siu | Kennedy Institute of Rheumatology, Nuffield Department of Orthopaedics, Rheumatology and Musculoskeletal Sciences, University of Oxford, Oxford, UK |
| Timothy J. Triche, Jr. | Department of Epigenetics, Van Andel Institute, Grand Rapids, MI, US |
| Alex Tsankov | Icahn School of Medicine at Mount Sinai |
| Sander W. van der Laan | University Medical Center Utrecht |
| Masanao Yajima | Boston University |
| Jean Yang | School of Mathematics and Statistics, Faculty of Science, University of Sydney, NSW 2006, NSW, Australia<br>Sydney Precision Data Science Centre, University of Sydney, Sydney, NSW 2006, Australia<br>Charles Perkins Centre, University of Sydney, Sydney, NSW 2006, Australia<br>Laboratory of Data Discovery for Health Limited (D24H), Science Park, Hong Kong SAR, China |
| Fabio Zanini | University of New South Wales |
| Ivana Jelic | Chan Zuckerberg Initiative |

## Author Contributions

Conceptualization: Ivana Jelic
Writing: Martin Hemberg, Federico Marini, Shila Ghazanfar, with input from all other authors.

# References

Aherrahrou, Rédouane, Dillon Lue, R. Noah Perry, Yonathan Tamrat Aberra, Mohammad Daud Khan, Joon Yuhl Soh, Tiit Örd, et al. 2023. "Genetic Regulation of SMC Gene Expression and Splicing Predict Causal CAD Genes." *Circulation Research* 132 (3): 323–38.

Aivazidis, Alexander, Fani Memi, Vitalii Kleshchevnikov, Brian Clarke, Oliver Stegle, and Omer Bayraktar. 2023. "Model-Based Inference of RNA Velocity Modules Improves Cell Fate Prediction." *bioRxiv*. https://doi.org/10.1101/2023.08.03.551650.

Brazma, A., P. Hingamp, J. Quackenbush, G. Sherlock, P. Spellman, C. Stoeckert, J. Aach, et al. 2001. "Minimum Information about a Microarray Experiment (MIAME)-toward Standards for Microarray Data." *Nature Genetics* 29 (4): 365–71.

Cannon, Matthew, James Stevenson, Kathryn Stahl, Rohit Basu, Adam Coffman, Susanna Kiwala, Joshua F. McMichael, et al. 2024. "DGIdb 5.0: Rebuilding the Drug-Gene Interaction Database for Precision Medicine and Drug Discovery Platforms." *Nucleic Acids Research* 52 (D1): D1227–35.

Cao, Yue, Yingxin Lin, John T. Ormerod, Pengyi Yang, Jean Y. H. Yang, and Kitty K. Lo. 2019. "scDC: Single Cell Differential Composition Analysis." *BMC Bioinformatics* 20 (Suppl 19): 721.

Cao, Yue, Lijia Yu, Marni Torkel, Sanghyun Kim, Yingxin Lin, Pengyi Yang, Terence P. Speed, Shila Ghazanfar, and Jean Yee Hwa Yang. 2023. "The Current Landscape and Emerging Challenges of Benchmarking Single-Cell Methods." *bioRxiv*. https://doi.org/10.1101/2023.12.19.572303.

Chen, Zhifen, and Heribert Schunkert. 2021. "Genetics of Coronary Artery Disease in the Post-GWAS Era." *Journal of Internal Medicine* 290 (5): 980–92.

Chu, Yanshuo, Enyu Dai, Yating Li, Guangchun Han, Guangsheng Pei, Davis R. Ingram, Krupa Thakkar, et al. 2023. "Pan-Cancer T Cell Atlas Links a Cellular Stress Response State to Immunotherapy Resistance." *Nature Medicine* 29 (6): 1550–62.

Cock, Peter J. A., Christopher J. Fields, Naohisa Goto, Michael L. Heuer, and Peter M. Rice. 2010. "The Sanger FASTQ File Format for Sequences with Quality Scores, and the Solexa/Illumina FASTQ Variants." *Nucleic Acids Research* 38 (6): 1767–71.

Crowell, Helena L., Charlotte Soneson, Pierre-Luc Germain, Daniela Calini, Ludovic Collin, Catarina Raposo, Dheeraj Malhotra, and Mark D. Robinson. 2020a. "Muscat Detects Subpopulation-Specific State Transitions from Multi-Sample Multi-Condition Single-Cell Transcriptomics Data." *Nature Communications* 11 (1): 6077.

———. 2020b. "Muscat Detects Subpopulation-Specific State Transitions from Multi-Sample Multi-Condition Single-Cell Transcriptomics Data." *Nature Communications* 11 (1): 6077.

Cui, Haotian, Chloe Wang, Hassaan Maan, Kuan Pang, Fengning Luo, Nan Duan, and Bo Wang. 2024. "scGPT: Toward Building a Foundation Model for Single-Cell Multi-Omics Using Generative AI." *Nature Methods*, February. https://doi.org/10.1038/s41592-024-02201-0.

CZI Single-Cell Biology Program, Shibla Abdulla, Brian Aevermann, Pedro Assis, Seve Badajoz, Sidney M. Bell, Emanuele Bezzi, et al. 2023a. "CZ CELL×GENE Discover: A Single-Cell Data Platform for Scalable Exploration, Analysis and Modeling of Aggregated Data." *bioRxiv*. https://doi.org/10.1101/2023.10.30.563174.

———. 2023b. "CZ CELL×GENE Discover: A Single-Cell Data Platform for Scalable Exploration, Analysis and Modeling of Aggregated Data." *bioRxiv*. https://doi.org/10.1101/2023.10.30.563174.

Dann, Emma, Neil C. Henderson, Sarah A. Teichmann, Michael D. Morgan, and John C. Marioni. 2022. "Differential Abundance Testing on Single-Cell Data Using K-Nearest Neighbor Graphs." *Nature Biotechnology* 40 (2): 245–53.

Diehl, Alexander D., Terrence F. Meehan, Yvonne M. Bradford, Matthew H. Brush, Wasila M. Dahdul, David S. Dougall, Yongqun He, et al. 2016. "The Cell Ontology 2016: Enhanced Content, Modularization, and Ontology Interoperability." *Journal of Biomedical Semantics* 7 (1): 44.

Dries, Ruben, Jiaji Chen, Natalie del Rossi, Mohammed Muzamil Khan, Adriana Sistig, and Guo-Cheng Yuan. 2021. "Advances in Spatial Transcriptomic Data Analysis." *Genome Research* 31 (10): 1706–18.

Emmanúel Antonsson, Sindri, and Páll Melsted. 2024. "Batch Correction Methods Used in Single Cell RNA-Sequencing Analyses Are Often Poorly Calibrated." *bioRxiv*. https://doi.org/10.1101/2024.03.19.585562.

Fischer, David S., Leander Dony, Martin König, Abdul Moeed, Luke Zappia, Lukas Heumos, Sophie Tritschler, Olle Holmberg, Hananeh Aliee, and Fabian J. Theis. 2021. "Sfaira Accelerates Data and Model Reuse in Single Cell Genomics." *Genome Biology* 22 (1): 248.

Grootaert, Mandy O. J., and Martin R. Bennett. 2021. "Vascular Smooth Muscle Cells in Atherosclerosis: Time for a Re-Assessment." *Cardiovascular Research* 117 (11): 2326–39.

Haniffa, Muzlifah, Deanne Taylor, Sten Linnarsson, Bruce J. Aronow, Gary D. Bader, Roger A. Barker, Pablo G. Camara, et al. 2021. "A Roadmap for the Human Developmental Cell Atlas." *Nature* 597 (7875): 196–205.
Han, Ya, Yuting Wang, Xin Dong, Dongqing Sun, Zhaoyang Liu, Jiali Yue, Haiyun Wang, Taiwen Li, and Chenfei Wang. 2023. "TISCH2: Expanded Datasets and New Tools for Single-Cell Transcriptome Analyses of the Tumor Microenvironment." *Nucleic Acids Research* 51 (D1): D1425–31.
Heumos, Lukas, Anna C. Schaar, Christopher Lance, Anastasia Litinetskaya, Felix Drost, Luke Zappia, Malte D. Lücken, et al. 2023. "Best Practices for Single-Cell Analysis across Modalities." *Nature Reviews. Genetics* 24 (8): 550–72.
Hicks, Stephanie C., Ruoxi Liu, Yuwei Ni, Elizabeth Purdom, and Davide Risso. 2021. "Mbkmeans: Fast Clustering for Single Cell Data Using Mini-Batch K-Means." *PLoS Computational Biology* 17 (1): e1008625.
Hie, Brian, Hyunghoon Cho, Benjamin DeMeo, Bryan Bryson, and Bonnie Berger. 2019. "Geometric Sketching Compactly Summarizes the Single-Cell Transcriptomic Landscape." *Cell Systems* 8 (6): 483–93.e7.
Hsieh, Chiao-Yu, Jian-Hung Wen, Shih-Ming Lin, Tzu-Yang Tseng, Jia-Hsin Huang, Hsuan-Cheng Huang, and Hsueh-Fen Juan. 2023. "scDrug: From Single-Cell RNA-Seq to Drug Response Prediction." *Computational and Structural Biotechnology Journal* 21: 150–57.
HuBMAP Consortium. 2019. "The Human Body at Cellular Resolution: The NIH Human Biomolecular Atlas Program." *Nature* 574 (7777): 187–92.
Jain, Sanjay, Liming Pei, Jeffrey M. Spraggins, Michael Angelo, James P. Carson, Nils Gehlenborg, Fiona Ginty, et al. 2023. "Advances and Prospects for the Human BioMolecular Atlas Program (HuBMAP)." *Nature Cell Biology* 25 (8): 1089–1100.
Kanemaru, Kazumasa, James Cranley, Daniele Muraro, Antonio M. A. Miranda, Siew Yen Ho, Anna Wilbrey-Clark, Jan Patrick Pett, et al. 2023. "Spatially Resolved Multiomics of Human Cardiac Niches." *Nature* 619 (7971): 801–10.
Knox, Craig, Mike Wilson, Christen M. Klinger, Mark Franklin, Eponine Oler, Alex Wilson, Allison Pon, et al. 2024. "DrugBank 6.0: The DrugBank Knowledgebase for 2024." *Nucleic Acids Research* 52 (D1): D1265–75.
Kuhn, Michael, Ivica Letunic, Lars Juhl Jensen, and Peer Bork. 2016. "The SIDER Database of Drugs and Side Effects." *Nucleic Acids Research* 44 (D1): D1075–79.
Lederer, Alex R., Maxine Leonardi, Lorenzo Talamanca, Antonio Herrera, Colas Droin, Irina Khven, Hugo J. F. Carvalho, et al. 2024. "Statistical Inference with a Manifold-Constrained RNA Velocity Model Uncovers Cell Cycle Speed Modulations." *bioRxiv : The Preprint Server for Biology*, January. https://doi.org/10.1101/2024.01.18.576093.
Lei, Wanyue, Mengqin Yuan, Min Long, Tao Zhang, Yu-E Huang, Haizhou Liu, and Wei Jiang. 2023. "scDR: Predicting Drug Response at Single-Cell Resolution." *Genes* 14 (2). https://doi.org/10.3390/genes14020268.
Li, Mengwei, Xiaomeng Zhang, Kok Siong Ang, Jingjing Ling, Raman Sethi, Nicole Yee Shin Lee, Florent Ginhoux, and Jinmiao Chen. 2022. "DISCO: A Database of Deeply Integrated Human Single-Cell Omics Data." *Nucleic Acids Research* 50 (D1): D596–602.
Luecken, Malte D., M. Büttner, K. Chaichoompu, A. Danese, M. Interlandi, M. F. Mueller, D. C. Strobl, et al. 2022. "Benchmarking Atlas-Level Data Integration in Single-Cell Genomics." *Nature Methods* 19 (1): 41–50.
Miller, Samuel A., Robert A. Policastro, Shruthi Sriramkumar, Tim Lai, Thomas D. Huntington, Christopher A. Ladaika, Daeho Kim, Chunhai Hao, Gabriel E. Zentner, and Heather M. O'Hagan. 2021. "LSD1 and Aberrant DNA Methylation Mediate Persistence of Enteroendocrine Progenitors That Support -Mutant Colorectal Cancer." *Cancer Research* 81 (14): 3791–3805.
Moreno, Pablo, Silvie Fexova, Nancy George, Jonathan R. Manning, Zhichiao Miao, Suhaib Mohammed, Alfonso Muñoz-Pomer, et al. 2022. "Expression Atlas Update: Gene and Protein Expression in Multiple Species." *Nucleic Acids Research* 50 (D1): D129–40.
Moreno, Pablo, Ni Huang, Jonathan R. Manning, Suhaib Mohammed, Andrey Solovyev, Krzysztof Polanski, Wendi Bacon, et al. 2021. "User-Friendly, Scalable Tools and Workflows for Single-Cell RNA-Seq Analysis." *Nature Methods* 18 (4): 327–28.
Mullan, Kerry Alyce, My Ha, Sebastiaan Valkiers, Benson Ogunjimi, Kris Laukens, and Pieter Meysman. 2023. "STEGO.R: An Application to Aid in scRNA-Seq and scTCR-Seq Processing and Analysis." *bioRxiv*. https://doi.org/10.1101/2023.09.27.559702.
Osumi-Sutherland, David, Chuan Xu, Maria Keays, Adam P. Levine, Peter V. Kharchenko, Aviv Regev, Ed Lein, and Sarah A. Teichmann. 2021. "Cell Type Ontologies of the Human Cell Atlas." *Nature*

*Cell Biology* 23 (11): 1129–35.

Pan, Lu, Paolo Parini, Roman Tremmel, Joseph Loscalzo, Volker M. Lauschke, Bradley A. Maron, Paola Paci, et al. 2024. "Single Cell Atlas: A Single-Cell Multi-Omics Human Cell Encyclopedia." *Genome Biology* 25 (1): 104.

Phipson, Belinda, Choon Boon Sim, Enzo R. Porrello, Alex W. Hewitt, Joseph Powell, and Alicia Oshlack. 2022. "Propeller: Testing for Differences in Cell Type Proportions in Single Cell Data." *Bioinformatics* 38 (20): 4720–26.

Quake, Stephen R. 2022. "A Decade of Molecular Cell Atlases." *Trends in Genetics: TIG* 38 (8): 805–10.

Regev, Aviv, Sarah A. Teichmann, Eric S. Lander, Ido Amit, Christophe Benoist, Ewan Birney, Bernd Bodenmiller, et al. 2017. "The Human Cell Atlas." *eLife* 6 (December). https://doi.org/10.7554/eLife.27041.

Ren, Xianwen, Wen Wen, Xiaoying Fan, Wenhong Hou, Bin Su, Pengfei Cai, Jiesheng Li, et al. 2021. "COVID-19 Immune Features Revealed by a Large-Scale Single-Cell Transcriptome Atlas." *Cell* 184 (7): 1895–1913.e19.

Rood, Jennifer E., Aidan Maartens, Anna Hupalowska, Sarah A. Teichmann, and Aviv Regev. 2022a. "Impact of the Human Cell Atlas on Medicine." *Nature Medicine* 28 (12): 2486–96.

———. 2022b. "Impact of the Human Cell Atlas on Medicine." *Nature Medicine* 28 (12): 2486–96.

Rue-Albrecht, Kevin, Federico Marini, Charlotte Soneson, and Aaron T. L. Lun. 2018. "iSEE: Interactive SummarizedExperiment Explorer." *F1000Research* 7 (June): 741.

Ruiter, Skyler, Seth Wolfgang, Marc Tunnell, Timothy Triche Jr, Erin Carrier, and Zachary DeBruine. 2023. "Value-Compressed Sparse Column (VCSC): Sparse Matrix Storage for Redundant Data." https://doi.org/10.48550/ARXIV.2309.04355.

Schurch, Nicholas J., Pietá Schofield, Marek Gierliński, Christian Cole, Alexander Sherstnev, Vijender Singh, Nicola Wrobel, et al. 2016. "How Many Biological Replicates Are Needed in an RNA-Seq Experiment and Which Differential Expression Tool Should You Use?" *RNA* 22 (6): 839–51.

Singh, Param Priya, and Bérénice A. Benayoun. 2023. "Considerations for Reproducible Omics in Aging Research." *Nature Aging* 3 (8): 921–30.

Song, Dongyuan, Qingyang Wang, Guanao Yan, Tianyang Liu, Tianyi Sun, and Jingyi Jessica Li. 2024. "scDesign3 Generates Realistic in Silico Data for Multimodal Single-Cell and Spatial Omics." *Nature Biotechnology* 42 (2): 247–52.

Song, Dongyuan, Nan Miles Xi, Jingyi Jessica Li, and Lin Wang. 2022. "scSampler: Fast Diversity-Preserving Subsampling of Large-Scale Single-Cell Transcriptomic Data." *Bioinformatics* 38 (11): 3126–27.

Squair, Jordan W., Matthieu Gautier, Claudia Kathe, Mark A. Anderson, Nicholas D. James, Thomas H. Hutson, Rémi Hudelle, et al. 2021a. "Confronting False Discoveries in Single-Cell Differential Expression." *Nature Communications* 12 (1): 5692.

———. 2021b. "Confronting False Discoveries in Single-Cell Differential Expression." *Nature Communications* 12 (1): 5692.

Subramanian, Aravind, Rajiv Narayan, Steven M. Corsello, David D. Peck, Ted E. Natoli, Xiaodong Lu, Joshua Gould, et al. 2017. "A Next Generation Connectivity Map: L1000 Platform and the First 1,000,000 Profiles." *Cell* 171 (6): 1437–52.e17.

Svensson, Valentine, Eduardo da Veiga Beltrame, and Lior Pachter. 2020. "A Curated Database Reveals Trends in Single-Cell Transcriptomics." *Database: The Journal of Biological Databases and Curation* 2020 (November). https://doi.org/10.1093/database/baaa073.

Svensson, Valentine, Roser Vento-Tormo, and Sarah A. Teichmann. 2018. "Exponential Scaling of Single-Cell RNA-Seq in the Past Decade." *Nature Protocols* 13 (4): 599–604.

Tarhan, Leyla, Jon Bistline, Jean Chang, Bryan Galloway, Emily Hanna, and Eric Weitz. 2023. "Single Cell Portal: An Interactive Home for Single-Cell Genomics Data." *bioRxiv : The Preprint Server for Biology*, July. https://doi.org/10.1101/2023.07.13.548886.

Theodoris, Christina V., Ling Xiao, Anant Chopra, Mark D. Chaffin, Zeina R. Al Sayed, Matthew C. Hill, Helene Mantineo, et al. 2023. "Transfer Learning Enables Predictions in Network Biology." *Nature* 618 (7965): 616–24.

Thomas, Tom, Charlotte Rich-Griffin, Mathilde Pohin, Matthias Friedrich, Dominik Aschenbrenner, Julia Pakpoor, Ashwin Jainarayanan, et al. 2023. "A Longitudinal Single-Cell Therapeutic Atlas of Anti-Tumour Necrosis Factor Treatment in Inflammatory Bowel Disease." *bioRxiv*. https://doi.org/10.1101/2023.05.05.539635.

Truong, Danh D., Salah-Eddine Lamhamedi-Cherradi, Robert W. Porter, Sandhya Krishnan, Jyothishmathi Swaminathan, Amber Gibson, Alexander J. Lazar, et al. 2023. "Dissociation Protocols Used for Sarcoma Tissues Bias the Transcriptome Observed in Single-Cell and Single-Nucleus RNA

Sequencing." *BMC Cancer* 23 (1): 488.

Vaidya, Anurag, Richard J. Chen, Drew F. K. Williamson, Andrew H. Song, Guillaume Jaume, Yuzhe Yang, Thomas Hartvigsen, et al. 2024. "Demographic Bias in Misdiagnosis by Computational Pathology Models." *Nature Medicine* 30 (4): 1174–90.

Van de Sande, Bram, Joon Sang Lee, Euphemia Mutasa-Gottgens, Bart Naughton, Wendi Bacon, Jonathan Manning, Yong Wang, et al. 2023. "Applications of Single-Cell RNA Sequencing in Drug Discovery and Development." *Nature Reviews. Drug Discovery* 22 (6): 496–520.

Wang, Yichen, Irzam Sarfraz, Wei Kheng Teh, Artem Sokolov, Brian R. Herb, Heather H. Creasy, Isaac Virshup, et al. 2023. "Matrix and Analysis Metadata Standards (MAMS) to Facilitate Harmonization and Reproducibility of Single-Cell Data." *bioRxiv : The Preprint Server for Biology*, March. https://doi.org/10.1101/2023.03.06.531314.

Wilkinson, Mark D., Michel Dumontier, I. Jsbrand Jan Aalbersberg, Gabrielle Appleton, Myles Axton, Arie Baak, Niklas Blomberg, et al. 2016. "The FAIR Guiding Principles for Scientific Data Management and Stewardship." *Scientific Data* 3 (March): 160018.

Williams, Cameron G., Hyun Jae Lee, Takahiro Asatsuma, Roser Vento-Tormo, and Ashraful Haque. 2022. "An Introduction to Spatial Transcriptomics for Biomedical Research." *Genome Medicine* 14 (1): 1–18.

Zdrazil, Barbara, Eloy Felix, Fiona Hunter, Emma J. Manners, James Blackshaw, Sybilla Corbett, Marleen de Veij, et al. 2024. "The ChEMBL Database in 2023: A Drug Discovery Platform Spanning Multiple Bioactivity Data Types and Time Periods." *Nucleic Acids Research* 52 (D1): D1180–92.

Zhang, Kai, Nathan R. Zemke, Ethan J. Armand, and Bing Ren. 2024. "A Fast, Scalable and Versatile Tool for Analysis of Single-Cell Omics Data." *Nature Methods*, January, 1–11.

Zhu, Chenxu, Sebastian Preissl, and Bing Ren. 2020. "Single-Cell Multimodal Omics: The Power of Many." *Nature Methods* 17 (1): 11–14.